# Characterizing quantum circuits with qubit functional configurations

Zixuan Hu and Sabre Kais*

*Department of Chemistry, Department of Physics, and Purdue Quantum Science and Engineering Institute, Purdue University, West Lafayette, IN 47907, United States*
*Email:* kais@purdue.edu

Abstract: We propose a theory of characterizing quantum circuits with qubit functional configurations. Any quantum circuit can be decomposed into alternating sequences of 1-qubit unitary gates and CNOT gates. Each CNOT sequence prepares the current quantum state into a layer of qubit functional configuration to specify the rule for the next 1-qubit unitary sequence on how to collectively modify the state vector entries. All the functional configuration layers on a quantum circuit define its type which can include many other circuits sharing the same configuration layers. Studying the functional configuration types allows us to collectively characterize the properties and behaviors of many quantum circuits. We demonstrate the theory's application to the hardware-efficient ansatzes of variational quantum algorithms. For potential applications, the functional configuration theory may allow systematic understanding and development of quantum algorithms based on their functional configuration types.

## 1. Introduction

The development of quantum algorithms with efficient quantum circuits has been a central part of quantum computation, which has seen enormous progress in the last 30 years in both the theoretical and experimental fronts [1-13]. Over the years, numerous quantum algorithms have been developed targeting a great variety of applications: these include the phase estimation algorithm [14], Shor's factorization algorithm [15], the Harrow-Hassidim-Lloyd algorithm for linear systems [16], the hybrid classical-quantum algorithms [17-19], the quantum machine learning algorithms [20, 21], and quantum algorithms for open quantum dynamics [22-24]. Despite the success, the design of efficient quantum circuits to implement new quantum algorithms remains an accidental process, and a systematic way to understand how quantum circuits work may lead to improvements in existing algorithms and discovery of new ones. To systematically study quantum circuit complexity, we have previously developed a theory of characterizing quantum states with their creation complexity [25]. In this work, we shift our attention from quantum states to quantum circuits and develop a systematic framework of characterizing quantum circuits with qubit functional configurations. The qubit functionals are first introduced in our previous work [26] as objects living in the dual space of the linear space formed by the basis states of the qubit state space. In this work we relate the qubit functionals to how the state vector entries are collectively modified by quantum gates. In particular, any arbitrary quantum circuit can be decomposed into alternating sequences of 1-qubit unitary gates and CNOT gates. Each CNOT sequence prepares the current quantum state into a layer of qubit functional configuration to specify the rule for the next $U_k$ sequence on how to collectively modify the state vector entries. All the layers together form a sequence of functional configurations that defines a unique type of quantum circuits with great variety. Characterizing quantum circuits by these functional configuration types has several major benefits: 1. all quantum circuits can be characterized by the corresponding functional configuration types; 2. each type characterizes important properties (such as circuit complexity) of the quantum circuits belonging to it; 3. each type contains a huge collection of possible quantum circuits allowing systematic investigation of their common properties. We demonstrate an application of the theory to the hardware-efficient ansatzes of variational quantum algorithms [27-30] by comparing the qubit functional configurations of two ansatzes used in Refs. [29] and [30]. We also propose a systematic process of creating any arbitrary functional configuration with the help of ancilla qubits. For potential applications, the functional configuration picture may allow us to systematically find a minimal gate decomposition sequence of any given quantum circuit. In addition, the classification may allow systematic understanding of quantum algorithms based on their quantum circuit types, which may benefit the development of new algorithms.

## 2. Theory of the qubit functional configuration

**2.1 The qubit functionals.** An *n*-qubit quantum state vector can be written as $|\varphi\rangle = \sum_{i=0}^{2^n-1} a_i |i\rangle$, where the basis states $|i\rangle$ can be associated with the bit strings $(i)$ with $i$ in the binary form. For example a 3-qubit state vector has 8 basis states corresponding to 8 bit strings: $(000)$, $(001)$, $(010)$, …, $(111)$. As discussed in Ref. [26] the collection of all these bit strings can be considered as a 3-dimensional linear space $V$ over the binary field $\{0,1\}$ and any bit string can be expressed as a linear combination of the three basis vectors (100), (010), and (001): $\mathbf{v} = (q_1, q_2, q_3) = q_1(100) + q_2(010) + q_3(001)$, where the coefficients $q_i$'s take values 0 or 1, and the addition "+" is bit-wise addition modulo 2. Now by the theory of linear vector spaces, the dual space $V^*$ of $V$ is formed by the linear functionals over $V$: $f(\mathbf{v}) = f((q_1, q_2, q_3)) = g_1 q_1 \oplus g_2 q_2 \oplus g_3 q_3$. Here the three basis functionals are $f^{(1)}((q_1, q_2, q_3)) = q_1$, $f^{(2)}((q_1, q_2, q_3)) = q_2$, $f^{(3)}((q_1, q_2, q_3)) = q_3$, the coefficients $g_i$'s take values 0 or 1, and the "$\oplus$" is addition modulo 2. In Ref. [26] we have associated each functional with a "0" condition that specifies a half-set of the bit strings and studied the quantum condition space generated by these "0" conditions. In this work we call these functionals "qubit functionals" (because they are functionals on the qubit values) and focus on their roles in how the entries of the quantum state vectors are modified collectively by quantum gates.

**2.2 Roles of elementary gates in modifying state vector entries.** The universality of quantum circuits says that any unitary operation can be decomposed into a sequence of 1-qubit unitaries and CNOT gates. Consider an arbitrary $n$-qubit quantum state $|\varphi_1\rangle = \sum_{i=0}^{2^n-1} a_i |i\rangle$ as the starting point, if we apply 1-qubit unitaries $U_k$ only without CNOT gates, then at most one $U_k$ can be applied on each qubit $q_k$ without redundancy. This is because $U_k$'s on different $q_k$'s all commute, so if e.g. $U_1$ has already been applied to $q_1$, then any additional unitary $V_1$ applied to $q_1$ at any point after $U_1$ will be equivalent to a single gate $W_1 = V_1 U_1$ applied to $q_1$, thus applying more than one gates on the same qubit is redundant and can be reduced to just one gate. Next to examine the actual effects of 1-qubit unitaries, we study a 3-qubit state $|\varphi_1\rangle = \sum_{i=0}^{7} a_i |i\rangle$ without loss of generality:

$$\varphi_1 = \begin{matrix}000\\001\\010\\011\\100\\101\\110\\111\end{matrix}\begin{pmatrix}a_0\\a_1\\a_2\\a_3\\a_4\\a_5\\a_6\\a_7\end{pmatrix} \xrightarrow{} U_1\varphi_1 = \begin{pmatrix}u_1 a_0 + u_2 a_4\\u_1 a_1 + u_2 a_5\\u_1 a_2 + u_2 a_6\\u_1 a_3 + u_2 a_7\\u_2 a_0 - u_1 a_4\\u_2 a_1 - u_1 a_5\\u_2 a_2 - u_1 a_6\\u_2 a_3 - u_1 a_7\end{pmatrix} ; \ U_2\varphi_1 = \begin{pmatrix}u_1 a_0 + u_2 a_2\\u_1 a_1 + u_2 a_3\\u_2 a_0 - u_1 a_2\\u_2 a_1 - u_1 a_3\\u_1 a_4 + u_2 a_6\\u_1 a_5 + u_2 a_7\\u_2 a_4 - u_1 a_6\\u_2 a_5 - u_1 a_7\end{pmatrix} ; \ U_3\varphi_1 = \begin{pmatrix}u_1 a_0 + u_2 a_1\\u_2 a_0 - u_1 a_1\\u_1 a_2 + u_2 a_3\\u_2 a_2 - u_1 a_3\\u_1 a_4 + u_2 a_5\\u_2 a_4 - u_1 a_5\\u_1 a_6 + u_2 a_7\\u_2 a_6 - u_1 a_7\end{pmatrix} \quad (1)$$

where the 1-qubit unitary $U_k$ has the same basic form $U = \begin{pmatrix} u_1 & u_2 \\ u_2 & -u_1 \end{pmatrix}$ applied to the $k^{th}$ qubit $q_k$: that is $U_1 = U \otimes I \otimes I$, $U_2 = I \otimes U \otimes I$, and $U_3 = I \otimes I \otimes U$ ($u_1$, $u_2$ are assumed real for cleaner notations; this assumption does not cause any reasoning or results below to lose generality as compared to using complex numbers; $u_1^2 + u_2^2 = 1$). Here we see that $U_1$ pairs $a_0$ with $a_4$, $a_1$ with $a_5$, $a_2$ with $a_6$, and $a_3$ with $a_7$. When we focus on one pair e.g. $a_0$ and $a_4$, the effect of $U_1$ is equivalent to $U\begin{pmatrix}a_0\\a_4\end{pmatrix} = \begin{pmatrix}u_1 a_0 + u_2 a_4\\u_2 a_0 - u_1 a_4\end{pmatrix}$, thus we can say $a_0$ plays the role of "0" in this 2-dimensional subspace while $a_4$ plays the role of "1". Now we see the effect of $U_1$ is dividing the total space into four 2-dimensional subspaces and then modifying the entries in pairs of $[(a_0, a_4), (a_1, a_5), (a_2, a_6), (a_3, a_7)]$ where the 1st entry in each parenthesis is considered "0" and the 2nd entry is considered "1". Similarly, the effect of $U_2$ can be represented by $[(a_0, a_2), (a_1, a_3), (a_4, a_6), (a_5, a_7)]$, and that of $U_3$ can be represented by $[(a_0, a_1), (a_2, a_3), (a_4, a_5), (a_6, a_7)]$. Note that $U_1$, $U_2$, and $U_3$ may be applied simultaneously but they keep the individual effects when considering a single gate alone. Now for $U_1$, $[(a_0, a_4), (a_1, a_5), (a_2, a_6), (a_3, a_7)]$ means that we have separated the total space into two half-spaces spanned by basis states with $q_1 = 0$: $\{|000\rangle \sim a_0, |001\rangle \sim a_1, |010\rangle \sim a_2, |011\rangle \sim a_3\}$ versus $q_1 = 1$: $\{|100\rangle \sim a_4, |101\rangle \sim a_5, |110\rangle \sim a_6, |111\rangle \sim a_7\}$; we then pair each $q_1 = 0$ term with a unique $q_1 = 1$ term and mix them to produce e.g. $U\begin{pmatrix} a_0 \\ a_4 \end{pmatrix} = \begin{pmatrix} u_1 a_0 + u_2 a_4 \\ u_2 a_0 - u_1 a_4 \end{pmatrix}$ or $U\begin{pmatrix} a_1 \\ a_5 \end{pmatrix} = \begin{pmatrix} u_1 a_1 + u_2 a_5 \\ u_2 a_1 - u_1 a_5 \end{pmatrix}$. Finally putting all these pairs back to the proper locations in the total vector, we obtain the vector result of $U_1 \varphi_1$ as in Equation (1). Similarly, $U_2$ separates the total space into two half-spaces spanned by basis states with $q_2 = 0$ versus $q_2 = 1$, while $U_3$ separates into half-spaces with $q_3 = 0$ versus $q_3 = 1$.

Now what happens if we apply some CNOT gates before the 1-qubit $U_k$'s?

$$\varphi_1 = \begin{matrix} 000 \\ 001 \\ 010 \\ 011 \\ 100 \\ 101 \\ 110 \\ 111 \end{matrix} \begin{pmatrix} a_0 \\ a_1 \\ a_2 \\ a_3 \\ a_4 \\ a_5 \\ a_6 \\ a_7 \end{pmatrix} \xrightarrow{\text{CNOT}_{1\to2}} \begin{matrix} 000 \\ 001 \\ 010 \\ 011 \\ 100 \\ 101 \\ 110 \\ 111 \end{matrix} \begin{pmatrix} a_0 \\ a_1 \\ a_2 \\ a_3 \\ a_6 \\ a_7 \\ a_4 \\ a_5 \end{pmatrix} \xrightarrow{\text{CNOT}_{2\to3}} \varphi'_1 = \begin{matrix} 000 \\ 001 \\ 010 \\ 011 \\ 100 \\ 101 \\ 110 \\ 111 \end{matrix} \begin{pmatrix} a_0 \\ a_1 \\ a_3 \\ a_2 \\ a_6 \\ a_7 \\ a_5 \\ a_4 \end{pmatrix} \qquad (2)$$

In Equation (2) we apply $\text{CNOT}_{1\to2}$ and then $\text{CNOT}_{2\to3}$ to produce $\varphi'_1$, where $i \to j$ in the subscripts means $q_i$ is the control and $q_j$ is the target. By studying either the truth table of the CNOT gates or the results from previous works [26, 31, 32], the $\text{CNOT}_{1\to2}$ calculates $q_1 \oplus q_2$ and stores its value on $q_2$, and afterwards the $\text{CNOT}_{2\to3}$ calculates $q_1 \oplus q_2 \oplus q_3$ and stores its value on $q_3$. In other words, after $\text{CNOT}_{1\to2}$ and $\text{CNOT}_{2\to3}$ have been applied, the value on $q_2$ now represents the value of $q_1 \oplus q_2$, and the value on $q_3$ now represents the value of $q_1 \oplus q_2 \oplus q_3$, as compared to the state $\varphi_1$ before the two CNOT gates are applied. Since both $q_1 \oplus q_2$ and $q_1 \oplus q_2 \oplus q_3$ are linear functionals in $V^*$, we can understand the vector $\varphi'_1$ as having a functional configuration of $(f_1 = q_1, f_2 = q_1 \oplus q_2, f_3 = q_1 \oplus q_2 \oplus q_3)$ as compared to $\varphi_1$ having $(f_1 = q_1, f_2 = q_2, f_3 = q_3)$. Now by the same reasoning as the previous paragraph we predict that, if we apply 1-qubit unitaries to $\varphi'_1$ now, $U_1$ will still separate the total space into two half-spaces spanned by basis states with $q_1 = 0$ versus $q_1 = 1$, but exactly which "0" term is paired with which "1" term will be different. $U_2$ now will separate into two half-spaces with $q_1 \oplus q_2 = 0$ versus $q_1 \oplus q_2 = 1$, and $U_3$ will separate into two half-spaces with $q_1 \oplus q_2 \oplus q_3 = 0$ versus $q_1 \oplus q_2 \oplus q_3 = 1$. Indeed this prediction is verified by carrying out the algebra on $\varphi'_1$ in the same manner as in Equation (1):

$$\varphi'_1 = \begin{pmatrix} a_0 \\ a_1 \\ a_3 \\ a_2 \\ a_6 \\ a_7 \\ a_5 \\ a_4 \end{pmatrix} \longrightarrow U_1\varphi'_1 = \begin{pmatrix} u_1 a_0 + u_2 a_6 \\ u_1 a_1 + u_2 a_7 \\ u_1 a_3 + u_2 a_5 \\ u_1 a_2 + u_2 a_4 \\ u_2 a_0 - u_1 a_6 \\ u_2 a_1 - u_1 a_7 \\ u_2 a_3 - u_1 a_5 \\ u_2 a_2 - u_1 a_4 \end{pmatrix} \; ; \; U_2\varphi'_1 = \begin{pmatrix} u_1 a_0 + u_2 a_3 \\ u_1 a_1 + u_2 a_2 \\ u_2 a_0 - u_1 a_3 \\ u_2 a_1 - u_1 a_2 \\ u_1 a_6 + u_2 a_5 \\ u_1 a_7 + u_2 a_4 \\ u_2 a_6 - u_1 a_5 \\ u_2 a_7 - u_1 a_4 \end{pmatrix} \; ; \; U_3\varphi'_1 = \begin{pmatrix} u_1 a_0 + u_2 a_1 \\ u_2 a_0 - u_1 a_1 \\ u_1 a_3 + u_2 a_2 \\ u_2 a_3 - u_1 a_2 \\ u_1 a_6 + u_2 a_7 \\ u_2 a_6 - u_1 a_7 \\ u_1 a_5 + u_2 a_4 \\ u_2 a_5 - u_1 a_4 \end{pmatrix} \quad (3)$$

In Equation (3) we see the effect of $U_1$ is $[(a_0, a_6), (a_1, a_7), (a_2, a_4), (a_3, a_5)]$ with the same "0" entries and "1" entries but different pairing from the previous $[(a_0, a_4), (a_1, a_5), (a_2, a_6), (a_3, a_7)]$. On the other hand, the effect of $U_2$ is $[(a_0, a_3), (a_1, a_2), (a_6, a_5), (a_7, a_4)]$ where the "0" entries are now $\{a_0, a_1, a_6, a_7\}$ – these entries in the original $\varphi_1$ vector (not $\varphi'_1$) correspond to $\{|000\rangle \sim a_0, |001\rangle \sim a_1, |110\rangle \sim a_6, |111\rangle \sim a_7\}$ – so indeed for these entries $q_1 \oplus q_2 = 0$. Similarly the effect of $U_3$ is $[(a_0, a_1), (a_3, a_2), (a_6, a_7), (a_5, a_4)]$ where the "0" entries are $\{a_0, a_3, a_6, a_5\}$ which correspond to $q_1 \oplus q_2 \oplus q_3 = 0$ in the original $\varphi_1$ vector.

**2.3 Characterizing quantum circuits with functional configurations.** With these results we are now ready to introduce the idea of using qubit functional configurations to characterize quantum circuits. Starting with an arbitrary initial state $|\varphi_1\rangle = \sum_{i=0}^{7} a_i |i\rangle$, we can consider the qubits as functional holders with the initial qubit functional configuration of $(f_1 = q_1, f_2 = q_2, f_3 = q_3)$. As long as there is no CNOT gate applied, this functional configuration remains in place and specifies a unique rule for all possible 1-qubit unitaries: which entries are considered "0", which are considered "1", and how the "0" and "1" entries are paired for each $U_k$. As long as the configuration stays in place, a maximum number of $n = 3$ $U_k$'s can be applied with one unitary on each qubit, and any additional 1-qubit unitary would be redundant. After the last $U_k$ under this configuration has been applied and before the first CNOT happens, the state changes to $|\varphi_2\rangle = \sum_{i=0}^{7} b_i |i\rangle$ with new entries in the state vector. Now suppose a sequence of CNOT gates are applied without any 1-qubit unitary in between, right before the next $U_k$ happens, the entries are not modified but only shuffled, and this creates a new layer of functional configuration such as $(f_1 = q_1, f_2 = q_1 \oplus q_2, f_3 = q_1 \oplus q_2 \oplus q_3)$ discussed above. This second layer of functional configuration now in place specifies its own unique rule for all possible 1-qubit unitaries that happen after, before the next CNOT gate is applied. Repeating this process we can analyze any arbitrary quantum circuits as follows:

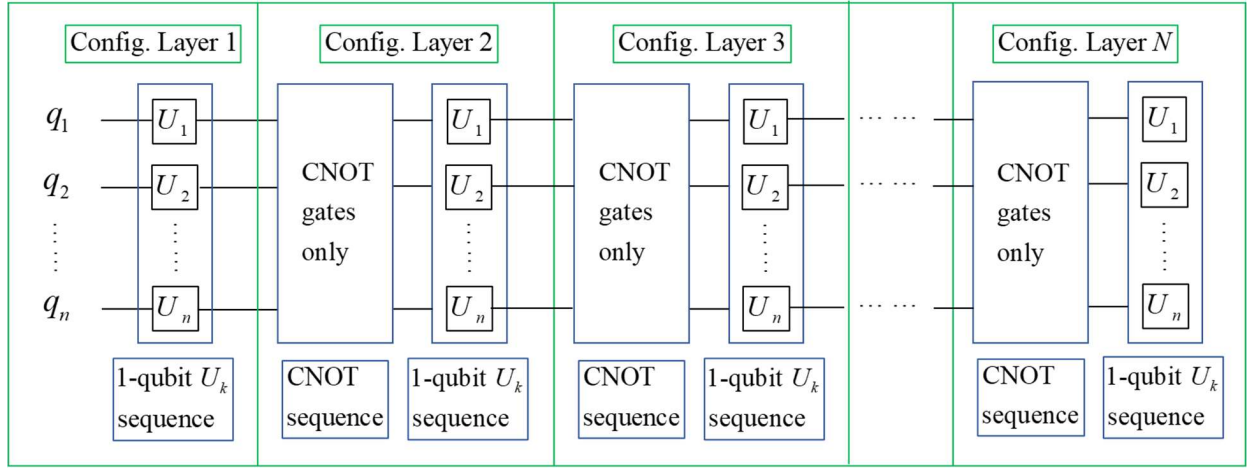

Figure 1. Graphical illustration of an arbitrary quantum circuit analyzed with multiple layers of qubit functional configurations. Note here we use the symbol $U_k$ to mean "any 1-qubit gate on the $k^{th}$ qubit", thus $U_k$ is not a particular gate and may represent different 1-qubit gates across different layers.

In Figure 1 an arbitrary quantum circuit can be decomposed into alternating sequences of 1-qubit $U_k$ and CNOT gates. Each CNOT sequence prepares the current quantum state into a new layer of qubit functional configuration to specify the rule for the next $U_k$ sequence on how the total space should be separated into "0" entries and "1" entries and how the "0" and "1" entries are paired. In particular, any $U_k$ finds the appropriate "0" and "1" entries according to the $k^{th}$ functional $f_k$ and the pairing pattern is defined by the entire functional configuration. After this $U_k$ sequence is completed, right before the next CNOT sequence begins, a quantum state with new entries is created, which can then be used as the initial state for the next iteration. This process can be repeated many times until the end of the circuit where the last layer of functional configuration specifies the rule for the final $U_k$ sequence to create the final state. Together, all the layers of functional configurations form a sequence of configurations that allows us to define a type of quantum circuits.

Next we demonstrate the application of the functional configuration picture in Figure 1 to analyzing the hardware-efficient ansatzes that are widely used in variational quantum algorithms [27-30]. The hardware-efficient ansatzes commonly include many identical layers, and within each layer there are two sub-layers: one sub-layer of parameterized 1-qubit unitaries and one sub-layer of two-qubit entanglers. The two-qubit entanglers are usually all CNOT gates with no 1-qubit unitaries in between, such that each entangler sub-layer can be considered as a single layer of qubit functional configuration as described above – this makes the hardware-efficient ansatzes natural application candidates for the qubit functional configuration theory. In the following examples, the ansatzes used in Ref. [29] and Ref. [30] are shown in Figure 2 and Figure 3 respectively, and the corresponding qubit functional configurations are shown in Equations (4) and (5) respectively.

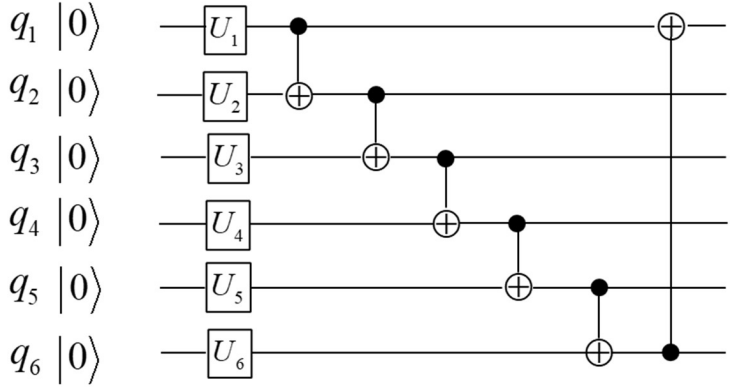

Figure 2. The quantum circuit of one layer of the hardware-efficient ansatz in Ref. [29].

The qubit functional configuration of the ansatz in Ref. [29]

$$f_1 = q_2 \oplus q_3 \oplus q_4 \oplus q_5 \oplus q_6, \; f_2 = q_1 \oplus q_2, \; f_3 = q_1 \oplus q_2 \oplus q_3, \; f_4 = q_1 \oplus q_2 \oplus q_3 \oplus q_4 \quad (4)$$
$$f_5 = q_1 \oplus q_2 \oplus q_3 \oplus q_4 \oplus q_5, \; f_6 = q_1 \oplus q_2 \oplus q_3 \oplus q_4 \oplus q_5 \oplus q_6$$

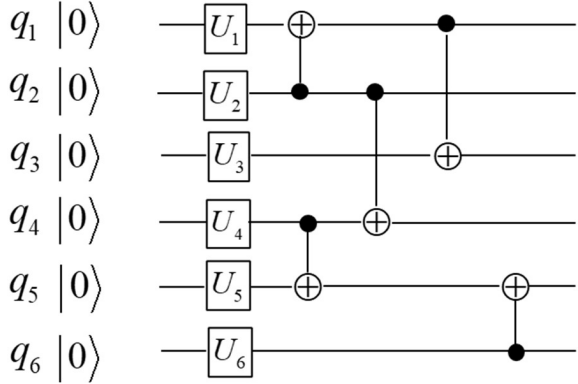

Figure 3. The quantum circuit of one layer of the hardware-efficient ansatz in Ref. [30].

The qubit functional configuration of the ansatz in Ref. [30]

$$f_1 = q_1 \oplus q_2, \; f_2 = q_2, \; f_3 = q_1 \oplus q_2 \oplus q_3, \; f_4 = q_2 \oplus q_4, \; f_5 = q_4 \oplus q_5 \oplus q_6, \; f_6 = q_6 \quad (5)$$

In Equations (4) and (5) the differences between the two ansatzes in Figure 2 and Figure 3 are clearly shown by the corresponding qubit functional configurations. For example, in Equation (5) $f_2$ and $f_6$ still keep their initial values $q_2$ and $q_6$, while in Equation (4) all the qubit functionals have been changed. In general, the qubit functionals in Equation (4) contain more qubit terms than those in Equation (5), thus we should expect the qubits of the circuit in Figure 2 to be more interconnected than those in Figure 3. This demonstrates the ability of the qubit functional configuration theory to characterize the properties of different ansatzes. In this particular case, the ansatzes will repeat the same functional configuration for all the following layers; however, if we apply the theory to other types of quantum circuits, such as the quantum Fourier transform circuit studied in the Supplementary Information (SI) Section S1, each layer may have a different functional configuration and the whole sequence of configuration layers characterize the circuit's properties.

Studying quantum circuits in terms of the qubit functional configuration types has various benefits:

1. Because the process in Figure 1 is entirely general, any arbitrary quantum circuit can be analyzed by this picture and thus belongs to one of the types defined by the functional configuration sequence.
2. As explained earlier, for any particular layer of functional configuration, a maximum number of $n$ 1-qubit unitaries can be applied without redundancy, with one unitary for each qubit. In addition, the number of CNOT gates required to achieve any given functional configuration is also bounded. Therefore the total length $L$ of a quantum circuit is closely related to the total number $N$ of functional configuration layers in the circuit. In other words, the functional configuration picture captures the important property of circuit complexity. The exact relation between $L$ and $N$ will be derived in Section 2.5.
3. More importantly, any particular sequence of the functional configurations contains the characteristic information of how the state vector entries are collectively modified by 1-qubit unitaries, while not fixing the actual $U_k$ sequences – this allows us to group a large collection of quantum circuits together to be classified under a single type defined by the functional configuration sequence. In fact, as the parameters of any 1-qubit unitary can be varied continuously in the complex number domain, there are uncountably infinite number of possible quantum circuits with a given length. However, for a functional configuration sequence of a given length, there are only finite number of possible sequences, thus by this picture we are able to classify an uncountably infinite collection with finite number of types.

Any particular type of quantum circuits has its unique sequence of functional configurations, and each layer of functional configuration defines a unique way by which 1-qubit unitaries work on the state vector entries. As detailed in Ref. [26], the duality between the linear functional space $V^*$ and the bit string space $V$ leads to a one-to-one correspondence between each functional and a unique partition of the bit strings – defined by two half-sets specified by setting "0" and "1" for the functional value respectively. When $U_k$ acts on $q_k$, the partition of $f_k$ specifies which entries are "0" and which are "1", while the partitions of all other functionals define how the "0" entries are paired with the "1" entries. Consequently if a particular functional $f_k$ appears in one configuration but not in another, then the 1-qubit unitaries will behave differently in the two configurations and thus two quantum circuits of different configuration types have different characteristics. Even if two configurations contain the same collection of functionals, two different permutations of these functionals will in general produce different quantum circuits. As detailed in the Supplementary Information (SI) Section S2, if there is only one layer, then e.g. the configuration $(f_1, f_2, f_3)$ can be considered somewhat equivalent to $(f_3, f_1, f_2)$ because $U_k$ in $(f_1, f_2, f_3)$ obeys the same rules as $U_{k+1}$ in $(f_3, f_1, f_2)$ with $U_{3+1} = U_1$. However, when multiple layers of configurations are present, then each corresponding configuration must permute in the same way for this equivalence to hold: e.g. $(f_1, f_2, f_3) \rightarrow (f_4, f_5, f_6)$ is equivalent to $(f_3, f_1, f_2) \rightarrow (f_6, f_4, f_5)$ but not to $(f_3, f_1, f_2) \rightarrow (f_4, f_6, f_5)$. Hence we can conclude that the types of quantum circuits as defined by different sequences of functional configurations do not in general overlap with each other. When studying properties of quantum circuits with the gate decomposition sequence one often encounters the problem of having multiple gate sequences that produce the same total unitary operation, which means that gate sequences do not uniquely characterize quantum circuits. In the functional configuration picture however, all these "equivalent" gate sequences must produce the same functional configuration sequence, and the types defined by the functional configuration sequences uniquely characterize quantum circuits. This is another benefit offered by the functional configuration picture. A detailed example of how different gate sequences can produce the same functional configuration is presented in the SI Section S3.

**2.4 Number of possible functional configuration types.** Now a natural question is to ask how many possible functional configuration types there are for a single layer and over a fixed number of layers. For a single layer, because the functional space is the dual of the bit string space with $2^n$ elements, there are $2^n$ functionals (for details see Ref. [26]). Excluding the "0" functional, there are $2^n - 1$ to choose from, and $n$ locations to allocate them. The most naïve count would be $(2^n - 1)^n$, but not all of these are valid configurations. To see this first note that it is impossible for duplicate functionals to appear in two or more different locations. This is because e.g. if we have a configuration of $(f_1, f_2 = f_1, f_3..., f_n)$ then the state vector will only have entries associated with the basis states $|00...\rangle$ and $|11...\rangle$ but not $|01...\rangle$ and $|10...\rangle$ (half of all dimensions have been lost for the state vector), and this cannot be always satisfied when the initial state are chosen arbitrarily. It turns out this non-duplication rule is a special example of a more general condition for valid configurations: for a configuration of $(f_1, f_2, f_3..., f_n)$, none of the $f_k$'s for $k = 1, 2, ..., n$ can be the sum of any number of other functionals. This is because otherwise if e.g. $f_7 = f_1 + f_2 + f_5$ then the value of $f_7$ is no longer free such that the subspace defined by $(f_1, f_2, f_5, f_7)$ becomes $(f_1, f_2, f_5)$ and half of the dimensions have been lost for the state vector. When we view the functionals as members of the linear space $V^*$, then requiring any $f_k$ is not the sum of any number of other functionals is the same as requiring linear independence for all $f_k$'s for $k = 1, 2, ..., n$ such that $\sum_{k=1}^{n} c_k f_k = 0$ has only the trivial solution of all $c_k$'s being 0. Note that this linear independence is automatically enforced by using CNOT gates to create functional configurations. To see this start with any configuration of $(f_1, f_2, f_3..., f_n)$ with all functionals linearly independent, then applying $CNOT_{j \rightarrow h}$ will replace $f_h$ with $f_j + f_h$ which is still linearly independent from all other functionals. So any CNOT gate cannot break the linear independence when starting with linear independence. The initial configuration of any quantum circuit is $(f_1 = q_1, f_2 = q_2, f_3 = q_3 ..., f_n = q_n)$ which has linear independence for all functionals, so indeed any CNOT sequence will automatically enforce the linear independence.

Now back to counting the possible configurations for a single layer, considering linear independence there are $2^n - 1$ choices for $f_1$, $2^n - 2$ for $f_2$, ..., for $f_k$ there are $N_f$ number of choices:

$$N_f(k,n) = 2^n - 1 - \sum_{i=1}^{k-1} C(k-1, i) = 2^n - 2^{k-1} \tag{6}$$

So the total possible number of configurations for one layer is:

$$N_c(n) = \prod_{k=1}^{n} N_f(k,n) = \prod_{k=1}^{n} \left(2^n - 2^{k-1}\right) \tag{7}$$

Now over multiple layers, for clarity we reinforce the definition of layers of functional configurations. Referring to Figure 1, start with the initial state $\varphi_1$ and the initial configuration $(f_1 = q_1, f_2 = q_2, f_3 = q_3 ..., f_n = q_n)$, if the first gate applied is a 1-qubit unitary (i.e. the case in Figure 1), we consider the initial configuration as the first layer; if the first gate is a CNOT gate (i.e. the case obtained by removing the first $U_k$ sequence and combining the first two blocks in Figure 1), we consider the configuration created by the first CNOT sequence as the first layer. After the first layer has been established, a sequence of 1-qubit unitaries can be applied according to the rule set by this configuration, and as long as there is no CNOT gate, the system stays in the first layer. Right before the next CNOT sequence happens, the state changes to $\varphi_2$, and the configuration is reset to $(f_1 = q_1, f_2 = q_2, f_3 = q_3 ..., f_n = q_n)$. Now the next CNOT sequence applies, and right before the next 1-qubit unitary sequence we create the configuration of the second layer, which stays in place until the next CNOT sequence. Subsequent layers can be defined accordingly. We require that all layers after the first one cannot be the initial configuration $(f_1 = q_1, f_2 = q_2, f_3 = q_3 ..., f_n = q_n)$ – because otherwise no CNOT gate is applied after the reset and we can combine this layer with the previous one – then the total possible number of types defined by $N$ layers of functional configurations is:

$$N_t(n, N) = \frac{N_c(n)\left(N_c(n) - 1\right)^{N-1}}{n!} = \frac{\prod_{k=1}^{n}\left(2^n - 2^{k-1}\right)\left[\prod_{k=1}^{n}\left(2^n - 2^{k-1}\right) - 1\right]^{N-1}}{n!} \tag{8}$$

where the $\left(N_c(n) - 1\right)^{N-1}$ accounts for "all layers after the first one cannot be the initial configuration", and $n!$ accounts for the equivalence due to permutation of functionals in the first layer. To get a concrete idea of $N_t$, for a 5-qubit system (like the simplest IBM quantum devices

[33] but with unlimited connectivity between qubits) over 1 layer of functional configuration, there are:

$$N_t(5,1) = \frac{\prod_{k=1}^{5}\left(2^5 - 2^{k-1}\right)}{5!} = 83328 \tag{9}$$

Over 3 layers of functional configurations, there are:

$$N_t(5,3) = \frac{\prod_{k=1}^{5}\left(2^5 - 2^{k-1}\right)\left[\prod_{k=1}^{5}\left(2^5 - 2^{k-1}\right) - 1\right]^2}{5!} = 8.3317318 \times 10^{18} \tag{10}$$

So the number of types $N_t$ can be a gigantic number even with quite small values for $n$ and $N$, but this is reasonable because we are considering all quantum circuits that are ever possible on $n$ qubits and $N$ layers with a finite number of types. In fact, the gigantic number $N_t$ illustrates the immense potential of quantum computation as a small number of qubits can support such enormous variety of quantum circuits over few layers.

**2.5 Circuit length in relation to the number of layers of functional configurations.** How does the number $N$ of layers relate to the length of the quantum circuit? As explained earlier, for any particular layer of functional configuration, a maximum number of $n$ 1-qubit unitaries can be applied without redundancy, with one unitary for each qubit. So for any quantum circuit with $N$ layers of functional configurations, the total number $N_U$ of 1-qubit unitaries is $N \leq N_U \leq nN$ (at least one unitary for each layer). This means that once the circuit type defined by the sequence of functional configurations has been fixed, the length of the circuit can only vary by up to $nN$. Now if more generally the circuit type is not fixed but the number of layers is fixed, to evaluate how many CNOT gates are possible for one layer, we need to study how many CNOT gates are required to reach an arbitrary functional configuration. Starting with any functional configuration $(f_1, f_2, f_3..., f_n)$, applying $\text{CNOT}_{j \to h}$ will replace $f_h$ with $f_j + f_h$ such that the process of creating functional configurations is dynamical and it is difficult to exactly evaluate how many CNOT gates are required. However, we can calculate an upper bound for the number of CNOT gates by storing the initial configuration of $(f_1 = q_1, f_2 = q_2, f_3 = q_3..., f_n = q_n)$ on $n$ ancilla qubits. To do this we start with the initial configuration $(f_1 = q_1, f_2 = q_2, f_3 = q_3..., f_n = q_n)$ on the computational qubits, prepare the ancilla qubits as $q_{a_1} = |0\rangle$, $q_{a_2} = |0\rangle$, ..., $q_{a_n} = |0\rangle$ and apply the gates $\text{CNOT}_{k \to a_k}$ for $k = 1,...,n$ with the computational qubits as controls and the ancilla qubits as targets, such to copy the configuration $\left(f_{a_1} = q_1, f_{a_2} = q_2, f_{a_3} = q_3..., f_{a_n} = q_n\right)$ on the ancilla qubits. This way when we modify the configuration on the computational qubits, the ancilla qubits still retain the initial configuration. After this we can create the wanted functionals one by one, e.g. if for the next layer the wanted configuration requires $f_1 = q_2 \oplus q_3 \oplus q_5$, then we can use the corresponding ancilla qubits as controls and the 1st computational qubit as target such as $\text{CNOT}_{a_1 \to 1}$, $\text{CNOT}_{a_2 \to 1}$, $\text{CNOT}_{a_3 \to 1}$, $\text{CNOT}_{a_5 \to 1}$ (order does not matter), and we will create $q_2 \oplus q_3 \oplus q_5$ on the 1st functional location. Note here $\text{CNOT}_{a_1 \to 1}$ is to remove $q_1$ from $f_1$. As the ancilla configuration is never changed in this process, we can repeat the same procedure when creating $f_2$, $f_3$, and all the way to $f_n$. This systematic process of creating any functional configuration is illustrated in Figure 4 with a 3-qubit example.

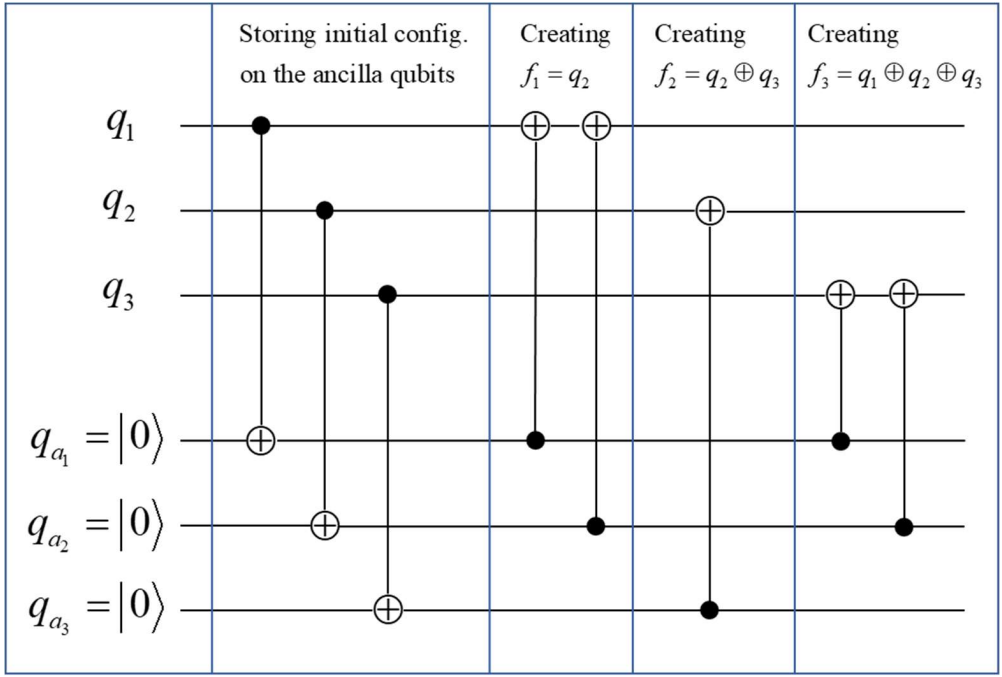

Figure 4. A 3-qubit example of systematically creating any functional configuration by first storing the initial configuration on 3 ancilla qubits and then creating the functionals one by one without changing the ancilla qubits.

With this process, for each $f_k$, the functional that requires the most CNOT gates is the sum of all qubits except $q_k$, e.g. $f_1 = q_2 \oplus q_3 \oplus ... \oplus q_n$, which requires $n$ CNOT gates. Therefore the number of CNOT gates required to reach any functional configuration cannot exceed $n \times n = n^2$. The situation is similar for all subsequent layers, hence the number of CNOT gates over $N$ layers cannot exceed $n^2 N$. In practice this upper bound may not be reached due to the linear independence constraint. Overall to achieve any single layer of configuration we need at most $n$ $\text{CNOT}_{k \to a_k}$ gates and $n^2$ $\text{CNOT}_{a_j \to k}$ gates, and the total over $N$ layers is $(n^2 + n)N$ CNOT gates. The least number we need is one CNOT for all the layers after the first one, so it is $(N-1)$ gates, and we have $N-1 \leq N_{\text{CNOT}} \leq (n^2 + n)N$, with the right equality not necessarily achievable. Now we see that even with just 5 qubits and a single layer $N=1$, the possible length of the quantum circuits is $1 \leq N_U \leq 5$ 1-qubit unitaries and $0 \leq N_{\text{CNOT}} \leq 30$ CNOT gates, so the number of types at 83328 in Equation (9) becomes more reasonable. Here we remark that on a case-by-case basis, the procedure to create functional configurations by ancilla qubits may be far less efficient compared to working with the computational qubits only. For example, the configuration $(f_1 = q_1, f_2 = q_1 \oplus q_2, f_3 = q_1 \oplus q_2 \oplus q_3)$ can be easily reached by $\text{CNOT}_{1 \to 2}$ and then $\text{CNOT}_{2 \to 3}$; while using ancilla qubits we first need three $\text{CNOT}_{k \to a_k}$ gates and then $\text{CNOT}_{a_1 \to 2}$, $\text{CNOT}_{a_1 \to 3}$, and $\text{CNOT}_{a_2 \to 3}$, requiring 6 CNOT gates in total. However the ancilla method is more systematic thus to allow us to bound the maximum number of CNOT gates needed. Even with this inefficiency, the upper bound of $nN$ and $(n^2 + n)N$ are both polynomial in $n$ and $N$ such that if $N$ is polynomial in $n$ then the maximum length of the quantum circuits classified by $N$ layers of functional configurations is also polynomial in $n$: this means that not only the quantum circuits belonging to the same type are either all polynomial or all exponential, but also the quantum circuits of types with the same number of layers are either all polynomial or all exponential. In other words, the functional configuration picture captures the important property of circuit complexity.

### 3. Potential applications

The qubit functional configuration picture described in Section 2 provides a systematic framework for characterizing quantum circuits. As discussed in Section 2.3, the framework assigns all possible quantum circuits of a given length into finite types that in general do not overlap. In addition, the characteristic information of how the state vector entries are collectively modified by 1-qubit unitaries is described by the functional configuration sequence. In the following we discuss potential applications provided by this new theory.

**3.1 Quantum algorithm development.** The classification of quantum circuits will allow us to systematically study circuits of the same type or similar types. Numerous quantum algorithms with diverse mechanisms and functionalities have been proposed in the past 30 years, yet these discoveries are mostly made on a case-by-case basis. A systematic study of these algorithms based on the functional configuration types will potentially help us understand why they work, and such knowledge may lead to improvements on existing algorithms and discovery of new algorithms. As demonstrated by the application examples in Section 2.3, The ansatzes used in variational quantum algorithms [27-30] are natural candidates for classification studies using the qubit functional configuration theory. The hybrid classical-quantum platform of the variational quantum algorithms has led to successful applications such as the variational quantum eigensolver (VQE) [17] and the quantum approximate optimization algorithm (QAOA) [19]. The properties and capabilities of variational quantum algorithms have been studied by descriptors such as expressibility and entangling capability [34] that describe how much of the solution space can be explored by varying the parameters. Now with the new qubit functional configuration picture we could classify various ansatzes into qubit functional configuration types such that their behaviors and properties are characterized by these types. In other words, the qubit functional configuration picture may allow us to systematically understand why certain ansatzes have better (or worse) performances when treating different kinds of problems.

**3.2 The minimal gate decomposition of a quantum circuit.** The qubit functional configuration picture may allow us to systematically find a minimal gate decomposition sequence of any given quantum circuit. Without the picture, the gate decomposition of a given quantum circuit is not unique (see Section 2.3 and the SI Section S3), and there is no systematic approach to find the minimal gate sequence or verify that a known gate sequence is minimal. With the picture, for any given quantum circuit, if one gate decomposition sequence (non-minimal) is known, we will then find the functional configuration sequence that defines the type for the circuit. Now for each layer of functional configuration, the minimal CNOT gate sequence to create this configuration is a well-defined sequence. Joining the minimal CNOT sequence for each configuration layer together and inserting the appropriate 1-qubit unitaries, we will have the minimal overall gate sequence for the quantum circuit. A worked-out example of how this may work is found in the SI SectionS2. Even further, as in Equation (7) there are only finite number of possible functional configurations for a single layer, thus for small numbers *n* of qubits, we could even construct a complete dictionary of all possible functional configurations with the corresponding minimal gate sequences, such that the minimal gate sequence for any arbitrary quantum circuit can be retrieved by searching the functional configuration index.

## 4. Conclusion

In this work we have developed a theory of characterizing all quantum circuits with qubit functional configurations. In particular, an arbitrary quantum circuit can be decomposed into alternating sequences of 1-qubit unitary $U_k$ gates and CNOT gates. Each CNOT sequence prepares the current quantum state into a layer of qubit functional configuration to specify the rule for the next $U_k$ sequence on how to collectively modify the state vector entries. All the layers together form a sequence of functional configurations that defines a unique type of quantum circuits with great variety. For uncountably infinite possible quantum circuits, there are only finite number of functional configuration types for a given length, and this number has been evaluated in detail. More importantly, these types uniquely characterize properties of the corresponding quantum circuits. We have demonstrated an application of the theory to the hardware-efficient ansatzes used in variational quantum algorithms by analyzing them in terms of the functional configurations. We have also proposed a systematic procedure of creating any arbitrary functional configuration by storing the initial configuration on ancilla qubits. For potential applications, the functional configuration picture captures the important property of circuit complexity of any given quantum circuit and may allow us to define a minimal sequence for it. The functional configuration picture may also allow systematic understanding and development of quantum algorithms.

**Acknowledgement**

ZH and SK acknowledge funding by the U.S. Department of Energy (Office of Basic Energy Sciences) under Award No. DE-SC0019215.

# Supplementary information: Characterizing quantum circuits with qubit functional configurations

Zixuan Hu and Sabre Kais*

*Department of Chemistry, Department of Physics, and Purdue Quantum Science and Engineering Institute, Purdue University, West Lafayette, IN 47907, United States*
*Email:* kais@purdue.edu

**S1. The qubit functional configuration layers of the quantum Fourier transform circuit.**

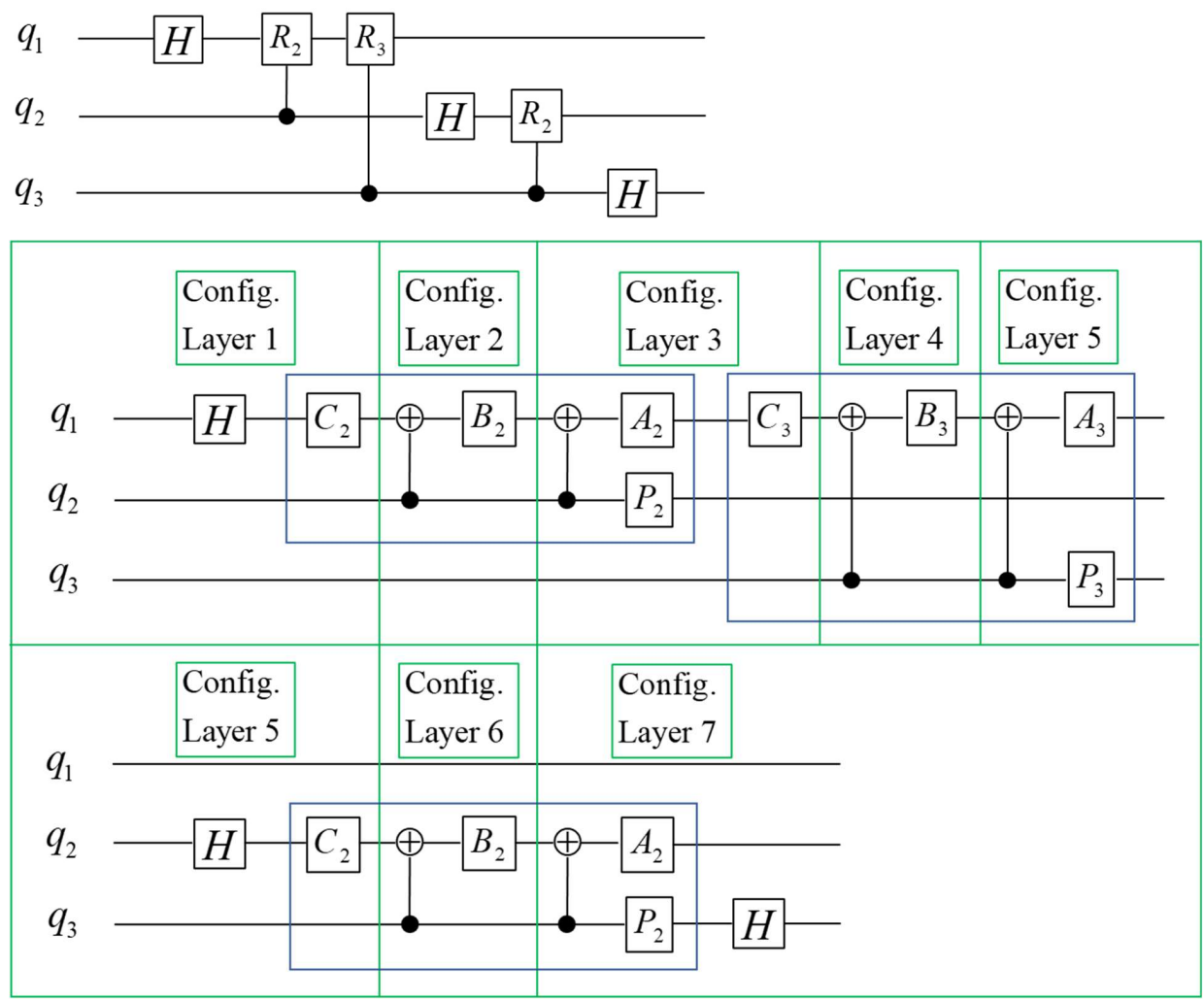

Figure S5. The 3-qubit quantum Fourier transform (QFT) circuit (top) analyzed with the functional configuration picture (bottom). The controlled-$R_h$ gates are decomposed into CNOT gates and 1-qubit unitaries as enclosed in the blue boxes (see the description below for the details of these gates). The functional configuration layers are indicated by the green boxes. There are 7 layers in total, with each layer after the 1st one defined by a single CNOT gate.

In the main text we have applied the functional configuration theory to the hardware-efficient ansatzes of variational quantum algorithms. For the ansatzes all the layers have the same functional configuration and hence we have only studied one layer. In this section we study the quantum Fourier transform (QFT) circuit that has different functional configurations on different layers, such that the circuit is described by a sequence of layers. The QFT [35] is an essential subroutine in many important quantum algorithms such as the phase estimation algorithm [14] and Shor's factorization algorithm [15]. Here we analyze the functional configuration sequence of a 3-qubit QFT circuit:

In Figure S5 we first decompose the controlled-$R_h$ gates, defined by $R_h = \begin{pmatrix} 1 & 0 \\ 0 & \exp(2\alpha_h i) \end{pmatrix}$ with $\alpha_h = \frac{2\pi}{2^{h+1}}$, into CNOT gates and 1-qubit unitaries as enclosed in the blue boxes. The standard decomposition [35] has $A_h = R_z\left(\frac{2\pi i}{2^h}\right)$, $B_h = C_h = R_z\left(\frac{-2\pi i}{2^{h+1}}\right)$, and $P_h = \begin{pmatrix} 1 & 0 \\ 0 & \exp(i\alpha_h) \end{pmatrix}$. The decomposed circuit on the bottom of Figure S5 contains 7 functional configuration layers as indicated by the green boxes, and the functional configuration sequence is: $(q_1, q_2, q_3) \rightarrow (q_1 \oplus q_2, q_2, q_3) \rightarrow (q_1 \oplus q_2, q_2, q_3) \rightarrow (q_1 \oplus q_3, q_2, q_3) \rightarrow (q_1 \oplus q_3, q_2, q_3) \rightarrow (q_1, q_2 \oplus q_3, q_3) \rightarrow (q_1, q_2 \oplus q_3, q_3)$. In this particular circuit the 1st configuration layer is the initial configuration of $(q_1, q_2, q_3)$, while each configuration layer after the 1st one is defined by a single CNOT gate, such that the number of CNOT gates is equal to the number of layers minus one. Considering the standard result of having $\frac{n^2 - n}{2}$ controlled-$R_h$ gates for an $n$-qubit system [35], and each controlled-$R_h$ gate being decomposed into two CNOT gates, the total number of functional configuration layers is $n^2 - n + 1$ for an $n$-qubit QFT circuit. We see the number of layers scales as $O(n^2)$ and thus agrees with the well-known gate scaling of the QFT circuit [35] – as detailed in the main text Section 2.5, in general the functional configuration picture captures the important property of circuit complexity.

**S2. Equivalence between qubit functional configurations due to permutation of functionals.**

In the main text we mentioned that if there is only one layer of functional configuration, and two functional configurations contain two permutations of the same collection of functionals, then the two configurations can be considered as equivalent. Below we give a concrete example of a 3-qubit system. Consider the two functional configurations of $C_1 = (f_1 = q_1, f_2 = q_2, f_3 = q_3)$ and $C_2 = (f_1 = q_1, f_2 = q_3, f_3 = q_2)$. Clearly $C_2$ can be obtained from $C_1$ by swapping $f_2$ and $f_3$. By studying the effects of CNOT gates on modifying functional configurations, the swap between $f_2$ and $f_3$ can be achieved by applying sequentially CNOT$_{2\rightarrow 3}$, CNOT$_{3\rightarrow 2}$, and CNOT$_{2\rightarrow 3}$ to $C_1 = (f_1 = q_1, f_2 = q_2, f_3 = q_3)$. To see this is correct, the configuration after the 1st $\text{CNOT}_{2\to 3}$ is $(f_1 = q_1, f_2 = q_2, f_3 = q_2 \oplus q_3)$, after $\text{CNOT}_{3\to 2}$ is $(f_1 = q_1, f_2 = q_3, f_3 = q_2 \oplus q_3)$, and after the 2nd $\text{CNOT}_{2\to 3}$ is $C_2 = (f_1 = q_1, f_2 = q_3, f_3 = q_2)$. Now suppose the initial state vector is $|\varphi_1\rangle = \sum_{i=0}^{7} a_i |i\rangle$, then we have:

$$\varphi_1 = \begin{pmatrix} 000 & a_0 \\ 001 & a_1 \\ 010 & a_2 \\ 011 & a_3 \\ 100 & a_4 \\ 101 & a_5 \\ 110 & a_6 \\ 111 & a_7 \end{pmatrix} \xrightarrow{\text{CNOT}_{2\to 3}} \begin{pmatrix} 000 & a_0 \\ 001 & a_1 \\ 010 & a_3 \\ 011 & a_2 \\ 100 & a_4 \\ 101 & a_5 \\ 110 & a_7 \\ 111 & a_6 \end{pmatrix} \xrightarrow{\text{CNOT}_{3\to 2}} \begin{pmatrix} 000 & a_0 \\ 001 & a_2 \\ 010 & a_3 \\ 011 & a_1 \\ 100 & a_4 \\ 101 & a_6 \\ 110 & a_7 \\ 111 & a_5 \end{pmatrix} \xrightarrow{\text{CNOT}_{2\to 3}} \varphi'_1 = \begin{pmatrix} 000 & a_0 \\ 001 & a_2 \\ 010 & a_1 \\ 011 & a_3 \\ 100 & a_4 \\ 101 & a_6 \\ 110 & a_5 \\ 111 & a_7 \end{pmatrix} \quad \text{S(1)}$$

where the $\varphi_1$ vector is in the configuration of $C_1 = (f_1 = q_1, f_2 = q_2, f_3 = q_3)$ and the $\varphi'_1$ vector is in the configuration of $C_2 = (f_1 = q_1, f_2 = q_3, f_3 = q_2)$. Clearly, $\varphi'_1$ compared to $\varphi_1$ is just swapping the entry $a_1$ for $|001\rangle$ with $a_2$ for $|010\rangle$, and swapping $a_5$ for $|101\rangle$ with $a_6$ for $|110\rangle$ -- this is just swapping $q_2$ and $q_3$. This means that applying a 1-qubit unitary on $q_2$ (or $q_3$) for $\varphi_1$ is equivalent to applying the same unitary on $q_3$ (or $q_2$) for $\varphi'_1$, and therefore $C_1 = (f_1 = q_1, f_2 = q_2, f_3 = q_3)$ is equivalent to $C_2 = (f_1 = q_1, f_2 = q_3, f_3 = q_2)$ in the sense that they contain the quantum circuits that perform equivalent operations. Therefore swapping functionals in the configuration leads to equivalent configurations. Now by the theory of permutation groups, any arbitrary permutation can be realized by a series of swaps, then we conclude that any configuration generated by permuting the functionals in the original configuration is equivalent to the original as both configurations have equivalent rules for the 1-qubit unitaries up to a permutation. This equivalence leads to the $n!$ term on the denominator in Equation (6) of the main text.

### S3. Multiple different CNOT gate sequences producing the same functional configuration and the minimal gate sequence.

In the main text we have discussed the situation where multiple different CNOT gate sequences can produce the same functional configuration and thus the same total unitary operation. Here we present a concrete example and discuss how a minimal gate sequence can be defined with the functional configuration picture. Again consider a 3-qubit system, the configuration of $(f_1 = q_1 \oplus q_2 \oplus q_3, f_2 = q_1 \oplus q_2, f_3 = q_1)$ can be created from the initial configuration of $(f_1 = q_1, f_2 = q_2, f_3 = q_3)$ by the CNOT sequence of $(\text{CNOT}_{1\to 2}, \text{CNOT}_{2\to 3}, \text{CNOT}_{1\to 3}, \text{CNOT}_{3\to 1}, \text{CNOT}_{1\to 3})$. However the same configuration can also be created from the initial configuration by another CNOT sequence of $(\text{CNOT}_{2\to 3}, \text{CNOT}_{3\to 1}, \text{CNOT}_{1\to 2}, \text{CNOT}_{3\to 2}, \text{CNOT}_{1\to 3})$. We see that these two CNOT sequences are equivalent in the sense that they produce exactly the same functional configuration. In fact, there can be more equivalent CNOT sequences such as $(\text{CNOT}_{3\to 1}, \text{CNOT}_{1\to 2}, \text{CNOT}_{2\to 3}, \text{CNOT}_{2\to 1}, \text{CNOT}_{3\to 1}, \text{CNOT}_{1\to 2}, \text{CNOT}_{2\to 1}, \text{CNOT}_{1\to 2}, \text{CNOT}_{2\to 3}, \text{CNOT}_{3\to 2}, \text{CNOT}_{2\to 3})$, which is longer than the previous two but still produce the same functional configuration. A major benefit of the functional configuration picture is it allows us to see the equivalence of these different CNOT sequences with ease, because they all correspond to the same functional configuration. Unlike gate sequences, types defined by functional configuration sequences uniquely characterize the corresponding quantum circuits.

This fact also allows us to define and find a minimal gate sequence for any given quantum circuit, because among all the equivalent gate sequences for any functional configuration, there must be a sequence or multiple sequences with the minimal number of gate count, and this sequence(s) will be the minimal required to create a part of the quantum circuit involving such a functional configuration. In fact if a gate sequence decomposition (may not be the minimal) is known for an arbitrary quantum circuit, for example the one described above: $(\text{CNOT}_{3\to 1}, \text{CNOT}_{1\to 2}, \text{CNOT}_{2\to 3}, \text{CNOT}_{2\to 1}, \text{CNOT}_{3\to 1}, \text{CNOT}_{1\to 2}, \text{CNOT}_{2\to 1}, \text{CNOT}_{1\to 2}, \text{CNOT}_{2\to 3}, \text{CNOT}_{3\to 2}, \text{CNOT}_{2\to 3})$, we can obtain the functional configuration for each layer based on this sequence (only one layer for this example): $(f_1 = q_1 \oplus q_2 \oplus q_3, f_2 = q_1 \oplus q_2, f_3 = q_1)$, and then find the minimal sequence to generate the same functional configuration: $(\text{CNOT}_{1\to 2}, \text{CNOT}_{2\to 3}, \text{CNOT}_{1\to 3}, \text{CNOT}_{3\to 1}, \text{CNOT}_{1\to 3})$, and this will be a much shorter circuit that performs the same task. If there are multiple layers, joining the minimal sequence for each layer together and inserting the appropriate 1-qubit unitaries will then yield the overall minimal sequence for the quantum circuit. Therefore by the functional configuration picture we have a systematic way of defining and finding the minimal sequence of any quantum circuit.